\documentclass[aps, prl , reprint,superscriptaddress]{revtex4-2}
\usepackage{verbatim}
\usepackage{bbold}
\usepackage{amsmath}
\usepackage[english]{babel}
\usepackage{graphicx}
\usepackage{mathtools}
\usepackage{tikz}
\usepackage{hyperref} 
\usepackage{textcomp}% for the \textmu
\usepackage[normalem]{ulem}
\hypersetup{colorlinks=true, linkcolor=blue, citecolor=blue, urlcolor=blue}
\newcommand{\ket}[1]{|#1\rangle}
\newcommand{\bra}[1]{\langle#1|}

\begin{document}
\title{Controlling coherence between waveguide-coupled quantum dots}

\author{D. Hallett}
\email[]{d.hallett@sheffield.ac.uk}
\affiliation{School of Mathematical and Physical Sciences, University of Sheffield, Sheffield, S3 7RH, UK}
\author{J.~Wiercinski} 
\email[]{jehw2000@hw.ac.uk}
\affiliation{SUPA, Institute of Photonics and Quantum Sciences, Heriot-Watt University, Edinburgh EH14 4AS, United Kingdom}
\author{L.~Hallacy}
\author{S.~Sheldon}
\author{R.~Dost}
\author{N.~Martin}
\author{A.~Fenzl}
\affiliation{School of Mathematical and Physical Sciences, University of Sheffield, Sheffield, S3 7RH, UK}
\author{I.~Farrer}
\author{A.~K.~Verma}
\affiliation{School of Electrical and Electronic Engineering, University of Sheffield, Sheffield, S1 3JD, UK}
\author{M.~Cygorek}
\affiliation{Condensed Matter Theory, Department of Physics, TU Dortmund, 44227 Dortmund, Germany}
\author{E.~M.~Gauger}
\affiliation{SUPA, Institute of Photonics and Quantum Sciences, Heriot-Watt University, Edinburgh EH14 4AS, United Kingdom}
\author{M.~S.~Skolnick}
\author{L.~R.~Wilson}
\affiliation{School of Mathematical and Physical Sciences, University of Sheffield, Sheffield, S3 7RH, UK}

\date{\today}
\begin{abstract}
    %We report the measurement of collective emission from a pair of independently tuneable InAs quantum dots embedded in a nanophotonic waveguide. 
    We present a novel waveguide design that incorporates a split-diode structure, allowing independent electrical control of transition energies of multiple emitters over a wide range with minimal loss in waveguide coupling efficiency. We use this design to systematically map out the transition from superradiant to independent emission from two quantum dots. We perform both lifetime as well as Hanbury~Brown-Twiss measurements on the device, observing anti-dips in the photon coincidences indicating collective emission while at the same time observing a drop in lifetime around zero detuning, indicating superradiant behaviour. Performing both measurement types allows us to investigate detuning regions which show both superradiant rate enhancement and inter-emitter coherence, as well as regions in which correlations persist in the absence of rate enhancement. 
\end{abstract}
\maketitle
%\section{Introduction}\label{secIntroduction}
%
\textit{Introduction.~~}
Multiple indistinguishable quantum emitters can couple to an optical mode as a single entity~\cite{Dicke1954}, experiencing enhanced light-matter coupling~\cite{ PhysRevLett.115.063601} and creating strongly correlated emission~\cite{ Atom_SR_Theory_1}. 
Such coherent interactions between quantum emitters are vital for the scaling up of many quantum information schemes~\cite{ PhysRevLett.83.4204, Aspuru-Guzik2012}. 
Optical coupling between multiple quantum emitters enables the generation of light with non-classical photon statistics~\cite{ W_S_Liu_1991, PhysRevA.91.043814, Mlynek2014} entanglement between qubits~\cite{PhysRevLett.110.080502, PhysRevA.85.032327, Chiral_SR_1_2015, Chiral_SR_4_2019}, and advanced quantum photonic circuits~\cite{doi:10.1126/science.1173731, RevModPhys.90.031002}. Moreover, quantum enhancement of the light-matter interaction is the basis for superabsorbers for light harvesting and quantum batteries~\cite{quach_superabsorption_2022}, while the concomitant effect of subradiance can protect quantum memories against losses~\cite{PhysRevLett.129.120502,  Application_QMem}, and enable long-range exciton transfer~\cite{PRXQuantum.3.020354, PhysRevX.11.041003}.

\begin{figure*}[ht]
\centering
\begin{tikzpicture}
    \node[anchor=south west, inner sep=0] (image) at (0,0) 
        {\includegraphics[width=1\linewidth]{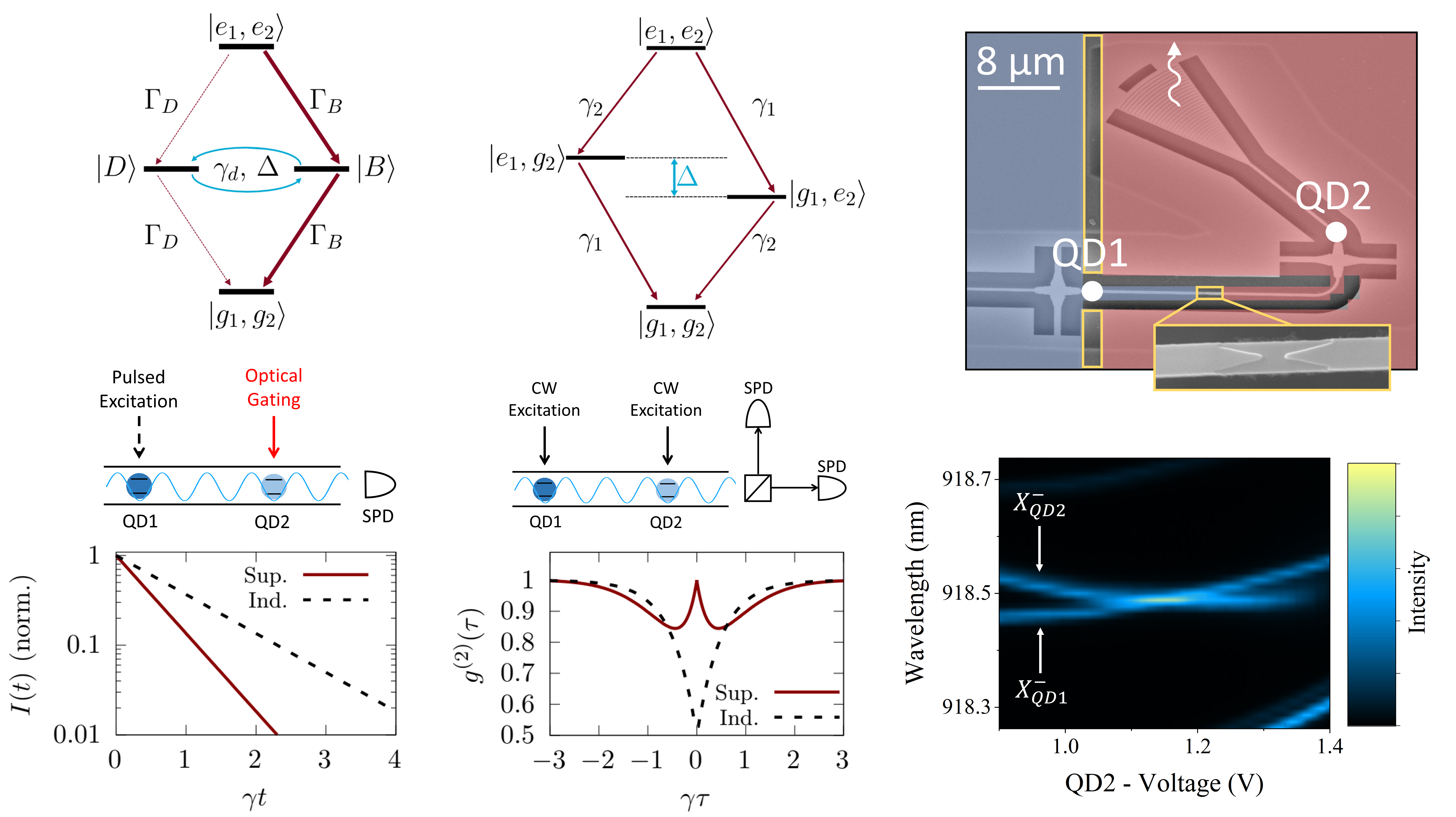}};
    \node at (0.50,9.5) [] {(a)};
    \node at (5.5,9.5) [] {(b)};
    \node at (11,9.5) [] {(e)};
    \node at (0.50,5.0) [] {(c)};
    \node at (5.5,5.0) [] {(d)};
    \node at (11,5.0) [] {(f)};

\end{tikzpicture}
\caption{\textbf{Independent tuning of QDs and properties of superradiant emission} (a) Energy level diagram of two resonant waveguide-coupled QDs. The states $\ket{e_1, e_2}$ and $\ket{g_1, g_2}$ describe both QDs in their excited and ground states, respectively. The spectral detuning and dephasing rate are labelled $\Delta$ and $\gamma_d$. (b) Energy level diagram of two independent quantum emitters. The state $\ket{e_1, g_2}$ ($\ket{g_1, e_2}$) signifies that $X^-_{QD1}$ ($X^-_{QD2}$) is excited, while the other QD is in the ground state. (c) Expected intensity decay for an ideal system of two superradiant emitters with identical decay rates $\gamma$ given the depicted excitation scheme. (d) Expected autocorrelation function $g^{(2)}(\tau)$ for a pair of superradiant emitters with identical decay rates $\gamma$ given the depicted excitation scheme. (e) SEM image of the waveguide device. The positions of the two QDs are marked by white spots. The red and blue colouring marks the area of the two diodes that control the two QDs. Yellow indicates the areas that are etched for electrical isolation. Inset – SEM of the isolation etch in the waveguide. (f) Emission spectrum of the $X^-$ transitions of the two QDs, measured from the outcoupler of the device, as a function of the voltage applied to QD2 (voltage applied to the red area in (e)).
}\label{fig1}
\end{figure*}

The most prominent effect of collective light matter coupling is superradiance, which emerges if an ensemble of identical emitters couples to the electromagnetic field. Superradiance manifests in a scaling of the emitted light intensity, $\langle{I}(t)\rangle\propto N^2$~\cite{Dicke1,gross_superradiance_1982}. For two emitters (cf.~Fig.~\ref{fig1}(a)), superradiance leads to enhanced transitions via a collective superposition state, the bright state $\ket{B}$, with rate $\Gamma_B$, while a dark state, $\ket{D}$, decouples from the electromagnetic field and decays with reduced rate $\Gamma_D$~\cite{QD_DR_3_2023, PhysRevResearch.6.033231}.Considering only decay from the single-excitation manifold, a linear increase in the emission rate is observed (cf.~Fig.~\ref{fig1}(c)) \cite{scheibner_superradiance_2007}. Introducing a detuning between the emitters leads to coherent oscillations between the dark and the bright state, until, if the detuning is significantly larger than the light-matter coupling, the system behaves as two individual emitters decaying with individual rates $\gamma_1$ and $\gamma_2$ (cf.~Fig.~\ref{fig1}(b)).

Additionally, Hanbury~Brown-Twiss (HBT) experiments allow measurement of inter-emitter coherences created by the transitions between delocalized states \cite{bhatti_superbunching_2015,Coop_Theory1_2023}. This leads to a characteristic anti-dip observed in the two-photon correlations around zero delay time (cf.~Fig.~\ref{fig1}(d)). 
Importantly, the anti-dip in the HBT signal results from relating the emission of a detected photon to a transition via a correlated emitter state (the bright state).
%For waveguide-coupled QDs this is naturally the case for photons at the waveguide out-coupler. 
%Therefore, the observation of an anti-dip does not by itself prove superradiance with its associated decay rate enhancement, but only confirms the presence of non-classical correlations in the emission process, which can be observed even for 
Such a transition can also be engineered for spatially separated, non-superradiant emitters via erasure of which-path information in the measurement process~\cite{doi:10.1021/acs.nanolett.6b03295, PhysRevLett.124.063603,QD_SR_3_2022,Coop_Theory1_2023,wiercinski_phonon_2023}.

%These applications motivate current efforts to demonstrate and control superradiance in solid-state quantum devices. 
Self-assembled quantum dots (QDs) have emerged as a leading platform for demonstrating and controling superradiance in solid-state quantum devices by virtue of combining the practicality of realising solid-state emitters at fixed positions with the excellent optical properties%of self-assembled QDs
~\cite{Zhai2020, Zhai2022}, and the 
integrability into nanophotonic devices%near-unity beta factor of nanophotonic waveguides
~\cite{lodahl_interfacing_2015, PhysRevLett.113.093603}. Nevertheless, there are several challenges to overcome in order to scale up superradiant solid-state quantum devices. First, QDs typically grow with random sizes and geometries, which 
makes it necessary to tune different QDs into resonance, for example 
%means that some way of tuning different QDs into resonance is required to achieve spectral indistinguishability of the QDs. Relative tuning of QD energies can be achieved 
through control of temperature~\cite{QD_SR_1_2018, doi:10.1021/acs.nanolett.6b03295}, strain~\cite{QD_SR_2_2019}, magnetic field~\cite{QD_DR_3_2023} or electric field~\cite{QD_SR_3_2022}. However, these tuning methods can be slow, irreversible, or unsuited for scale-up. Second, superradiance requires spatial indistiguishability of the quantum emitters.
An attractive solution to achieve superradiance between spatially separate solid-state emitters is therefore to coherently couple them to a photonic structure such as a waveguide~\cite{QD_SR_1_2018, QD_SR_2_2019}.
This facilitates an effective coupling to a set of confined one-dimensional modes, leading to the emergence of coherent QD-QD superposition states.% which couple to the electric field with enhanced (bright states) and reduced (dark state) decay rates (cf.~Fig.~\ref{fig1}C).

Existing experimental work on superradiance in solid-state emitters has used either the change in decay rate and the presence of a dark state~\cite{QD_DR_3_2023}, or photon coincidence measurements~\cite{QD_SR_1_2018, QD_SR_2_2019, SR_SV_1_2016} to demonstrate the presence of superradiance. For this, mostly the case of zero detuning has been compared to a far-detuned case, in which the emitters are assumed to behave independently. However, quantitative changes in the photon emission rate may have origins different from superradiance, e.g., variations of the dipole of a single emitter resulting from changes of the wave function upon tuning, or contributions from dark exciton states~\cite{lodahl_interfacing_2015}. An anti-dip observed in HBT experiments could in principle be used to indicate the presence of coherence between the QDs at resonance but not necessarily superradiance~\cite{QD_SR_3_2022}. %There, the anti-dip was caused by measurement-induced entanglement due to the erasure of which-path information about which QD emitted the first photon. 
Therefore, it is desirable to measure both photon coincidences as well as changes in the lifetime, while continuously varying the detuning between the QDs to demonstrate the presence of superradiance.  

Here, we demonstrate measurements of both lifetimes and photon coincidences on two independently tuneable QDs coupled to a nanobeam waveguide across a wide range of spectral detunings. Independent tuning of the QDs is achieved by separate electrical gating~\cite{PhysRevLett.131.033606}. This gating method allows fast, repeatable tuning of individual emitters, without compromising transmission through the photonic device, while at the same time being scalable to larger QD numbers.
Optical gating of one QD~\cite{PhysRevLett.108.057401, PhysRevB.87.115305} is used to switch between---and therefore directly compare---superradiant and non-superradiant regimes, by effectively adding and removing one emitter. Measurements of the decay rate of the system indicate a clear enhancement of the decay rate when the emitters are resonant. Additionally, we confirm the presence of strong QD-QD correlations by performing HBT experiments, even at detunings much larger than multiple single-emitter linewidths.
\\
\\
%%%%%%% New Section: Device design %%%%%%%%%%%%
\textit{Device design.~~}
The device studied in this work is pictured in Fig.~\ref{fig1}(e). Two InAs QDs are coupled to a nanobeam waveguide that supports a single guided TE mode. The nanobeam is 280~nm wide and 170~nm thick. The positions of the two QDs are marked by white circles. The distance between the QDs is approximately 20~\textmu m---roughly $70\times$ the wavelength of light in the waveguide. The two QDs can be separately addressed using two driving lasers, and emission is collected from the grating outcoupler at the end of the waveguide. 

In this work, we use separate Stark tuning of the two sections to control the detuning between the transitions of the QDs. To achieve this, the QDs are located in a split diode structure, allowing different voltages to be applied to different sections of the device. The membrane consists of a p-i-n diode with AlGaAs tunnelling barriers to increase the QD tuning range (see supplementary material \cite{SM}). A shallow etch is used to break electrical conductivity through the top p-doped layer of the diode, allowing different voltages to be applied to the separate regions. The positions of this etch are marked in yellow in Fig.~\ref{fig1}(e), and the red and blue colouring indicates the two electrically isolated regions. The geometry of the shallow etch through the waveguide is optimised using an inverse design technique to minimise loss; an SEM image of the shallow waveguide etch is inset in Fig.~\ref{fig1}(e). We estimate that the electrical isolation etch results in a $<1\%$ transmission loss across a broad bandwidth~\cite{SM}. At the same time, from numerical simulations we expect this waveguide design to achieve a coupling efficiency ($\beta$ factor) of around 0.8.

Relative tuning of the negatively charged trions of the two QDs ($X^-_{QD1, QD2}$) is shown in Fig.~\ref{fig1}(f). The trion states are identified by the lack of fine-structure splitting, and high turn-on voltage. For this measurement, both QDs are excited using above-band lasers and emission is collected from the outcoupler. The voltage applied to QD1 is held constant at 1.17~V. The voltage applied to QD2 (the red section in Fig.~\ref{fig1}(e) is tuned, blueshifting the emission from $X^-_{QD2}$ as the voltage is increased. It can be seen that $X^-_{QD1}$ is also tuned, due to cross-talk between the two diodes, with $X^-_{QD1}$ redshifting with increasing voltage. At approximately 1.15~V the two transitions are near-degenerate.
%%%%%%%%%%%%%%%%%%%%%%%%%%%%%%%%%%%%%%%%%%
\\
\\
\textit{Lifetime measurements~~}
\begin{figure*}[ht]
\centering
\begin{tikzpicture}
    \node[anchor=south west, inner sep=0] (image) at (0,0) 
        {\includegraphics[width=1\linewidth]{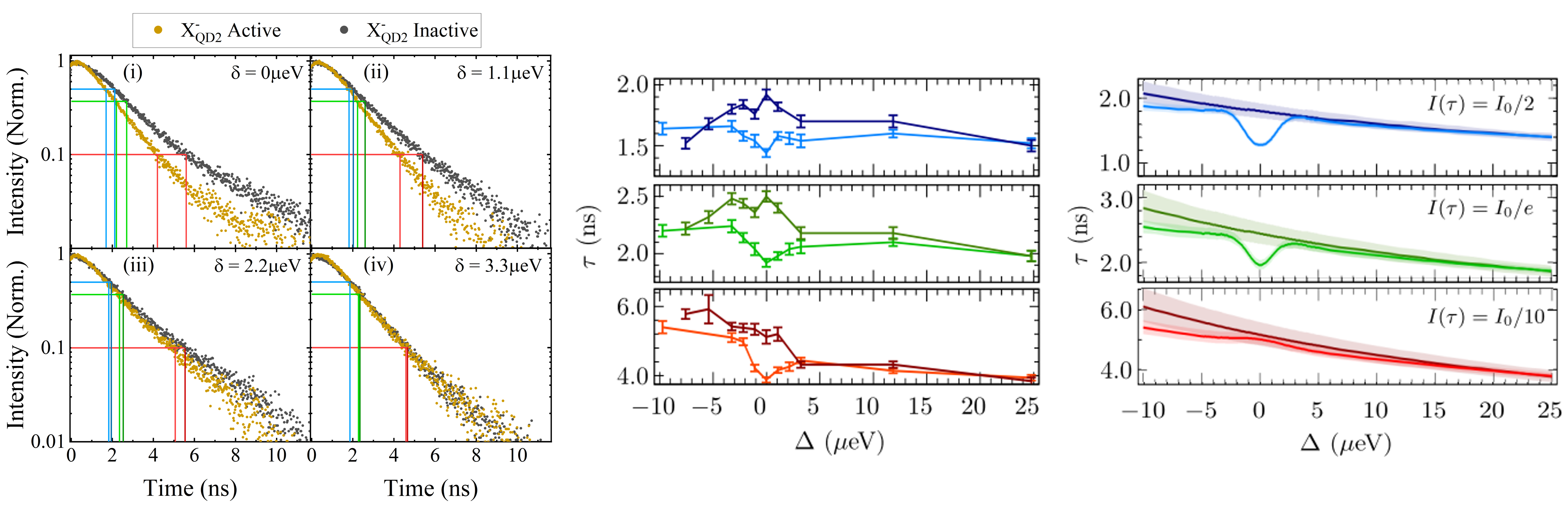}};
    \node at (0.05,5.5) [] {(a)};
    \node at (6.75,5.5) [] {(b)};
    \node at (12.4,5.5) [] {(c)};
\end{tikzpicture}
\caption{\textbf{Enhancement of emitter decay rate.} (a) Comparison of the decay curves of the system with the QDs detuned by (i) $0$ \textmu eV, (ii) 1.1 \textmu eV, (iii) 2.2 \textmu eV and (iv) 3.3 \textmu eV. Measurements are performed with (yellow) and without (black) optical gating of QD2. Coloured lines indicate where the intensity drops below the thresholds used in (b) and (c). 
(b) Detuning dependence of the decay rate with (lighter lines) and without (darker lines) gating of QD2. The results are plotted as the time taken for the emission intensity to drop below three threshold values: $I_0/2$ (blue), $I_0/e$ (green) and $I_0/10$ (red). (c) Theoretical prediction of the results in (b).
The shaded regions signify uncertainties in the model due to errors in the $X^-_{QD1}$ lifetime estimation (see main text).
}\label{fig2}
\end{figure*}
First, we explore how the radiative decay dynamics are affected by the detuning between the QDs using the excitation scheme shown in Fig.~\ref{fig1}(c). $X^-_{QD1}$ is excited with an above-band femtosecond pulse, emission is collected from the outcoupler and correlated with the time of the laser pulses to generate the emission decay curve. The detuning between the states is controlled by tuning the voltage applied to QD1. The effect of $X^-_{QD2}$ on the system is modulated through an optical gating effect~\cite{PhysRevLett.108.057401, PhysRevB.87.115305}. This effect controls the ground state of QD2, activating the negatively charged exciton transition, without populating the excited state. This gating is achieved using a low-power, continuous, above-band laser. Neither the pulsed laser nor the gating laser significantly excites the $X^-_{QD2}$ state. This allows us to explore the decay dynamics of a single QD as well as the coupled two QD system. 

The decay curves with and without optical gating of QD2 at different detunings are presented in Fig.~\ref{fig2}(a). From panel (i) It is apparent that the decay is faster with $X^-_{QD2}$ active and degenerate with $X^-_{QD1}$. This enhanced decay is very sensitive to detuning, and reduces continuously (ii-iii) until at detunings beyond just 3~{\textmu}eV (iv) the gating of QD2 no longer affects the decay rate. 
 
The decay of a pair of superradiant emitters is not expected to follow a simple exponential decay~\cite{QD_DR_3_2023}. Therefore, we choose a metric to assess the decay times in the sample that does not assume mono-exponential decay. In particular, we consider the time $\tau(\epsilon)$ it takes for the intensity to decay from its maximum value $I_0$ to a threshold of $\epsilon I_0$, where we choose $\epsilon = 1/2, 1/e, 1/10$. We assign error bars to the extracted decay times by considering the error of extracting the correct value for $\tau(\epsilon)$ for a given data set ~\cite{SM}. 

The detuning dependence of the decay times to these three thresholds is shown in Fig.~\ref{fig2}(b). Considering first only the results for $X^-_{QD1}$ (shown by the darker lines), the decay gradually gets quicker as the QD it tuned to positive detunings. This is most clear in the 1/10 (red) threshold. We attribute this to a change in the QD dipole as it is tuned. There is also a slight rise of between 5 and 10\% in the decay times for some thresholds around $\Delta = 0$. Note that at short timescales, these results are impacted by a timing jitter introduced by the non-resonant excitation. 

Comparing this to the case where $X^-_{QD2}$ is active (lighter lines in Fig.~\ref{fig2}(b)), we see that $X^-_{QD2}$ has little effect on the decay dynamics when the states are detuned. At resonance however, we see a significant reduction of about 20\% in decay time. We attribute this to a superradiant enhancement of the bright state decay rate $\Gamma_B$ with respect to the single QD decay rate $\gamma_1$. Ideal superradiance would lead to a reduction of decay time by a factor of 2. This is not observed in this sample due to a combination  of spectral wandering, a non-unity beta-factor, and the presence of other dephasing sources, e.g.~phonons~\cite{SM, PhysRevResearch.6.033231}. 

For comparison, we develop a theoretical model which considers a finite waveguide quality, decoherence and the non-resonant excitation of QD1~\cite{SM}.  As expected, for short to medium time scales our model predicts a drop in the decay time of the system (See Fig.~\ref{fig2}(c)). At long times, the intensity is predicted to be dominated by the presence of the dark state, leading to an increase in decay time (See Fig.~\ref{fig2}(c), lower panel). Note that this bi-exponential behaviour of the intensity has been seen in previous experiments~\cite{QD_DR_3_2023} but is not visible in our experimental data. 
We attribute this discrepancy to nonradiative decay processes which dominate over the small dark state decay rate. 
Overall, the lifetime measurements reveal that the emission behaviour of the sample is altered by optically activating $X^-_{QD2}$. These alterations are significant in the region around zero detuning and lead to a consistent reduction of the observed decay times, indicating a superradiant enhancement. 
%%%%%%%%%%%%%%%%%%%%%%%%%%%%%%%%%%%%%%%%%%%%%%%
\\
\\
\textit{Photon coincidence measurements~~}
%\subsection{Inter-emitter coherence during the emission process}\label{Sec3}
\begin{figure}[ht]
\centering
\includegraphics[width=1\linewidth]{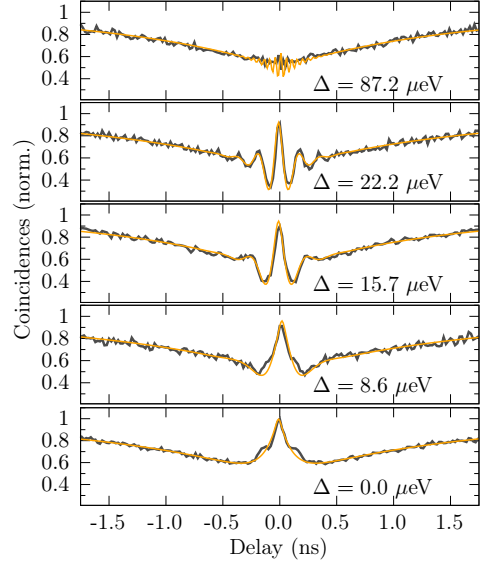}
\caption{\textbf{Photon correlations.} Autocorrelation function of emission collected from the end of the waveguide as a function of emitter-emitter detuning. The yellow lines are theoretical predictions of the measurements.}\label{fig3}
\end{figure}
In addition to the measurements of decay times presented in the previous section, HBT experiments were performed on the same pair of QDs. These measurements allow a direct observation of inter-emitter coherence during the emission process \cite{Coop_Theory1_2023, wiercinski_phonon_2023}. In this way, the presence of inter-emitter coherences can be validated as the source of the drop in radiative lifetime observed in the previous section. For this, both QDs are excited continuously with above-band CW lasers leading to incoherent population of both trions. Emission is collected from the out-coupler, spectrally filtered to isolate emission from the two QDs, split at a beam splitter, and sent to two single-photon detectors (SPDs). The detuning of the emitters is controlled by tuning the voltage across QD1. 

The results of the HBT experiments for different detunings can be seen in Fig.~\ref{fig3}. For large spectral detunings (top panel) the QDs show uncorrelated emission signified by a drop of the second-order autocorrelation function to 0.5 at zero time-delay. By contrast, when both QDs are tuned into resonance (bottom panel), an anti-dip approaching unity at zero delay time can be observed, which indicates the presence of inter-emitter coherence after the first photon emission, i.e., the preparation of the bright state by the emission process. At intermediate detunings (centre panels), coherence in the emission process can still be observed at detunings far larger than the single-emitter linewidth, and exceeding the region of rate enhancement that has been observed in lifetime measurements (see Fig.~\ref{fig2}(a)). The detuning between the emitters then manifests itself in a beating behaviour with a beating frequency corresponding exactly to the QD detuning~\cite{SM}. Thus, the transition between the correlated emission at zero detuning and uncorrelated emission can be explained by the beating frequency increasing, until it can no longer be resolved by the measurement.  
Using the measured spectral detuning as input, our theoretical model captures the transition between correlated and independent emission and reproduces the height and frequency of the beating with an estimate of spectral wandering of $\sigma=1.3$~\textmu {eV} and a pure dephasing rate of $\gamma_d=(8.0\pm1.6)~\textrm{ns}^{-1}$ which are identical across all HBT experiments~\cite{SM}. 
%We notice that our theoretical model does not reproduce the small shoulders in the anti-dip of the zero detuning data. The origin of these shoulder may lie in the spectral wandering of the QDs, which is likely to contain jumps that are not captured in the Gaussian line broadening model utilised here.
\\
\\
\textit{Discussion~~}
In this work we have presented a gated architecture combined with an optimised waveguide design that enables precise, independent control of emitter wavelengths while maintaining high coupling efficiencies to the waveguide optical modes. This control is fast, reversible and easily scalable. This design was used to create and control coherence between two QDs at a distance 70 times larger than the emission wavelength. %This architecture including a fast, reversible, and easily scalable tuning method is therefore an important step towards realising large-scale condensed-matter based quantum networks.

We have performed combined HBT and lifetime measurements on two QDs embedded in a waveguide device. Systematically tuning the QDs in and out of resonance, we demonstrated the transition from a superradiant system to fully independent emitters. At low to zero detuning we demonstrate a decrease in lifetime of about 20\%. We validate that this change in lifetime is due to a superradiant rate enhancement by confirming the simultaneous presence of an anti-dip in the HBT data, which indicates the presence of inter-emitter coherences.
%while at the same time observing an anti-dip around zero delay in the HBT data, which indicates the presence of inter-emitter coherences. 
Additionally, we investigate a region of intermediate detunings, where photon correlations show non-classical behaviour even at detunings of several single-emitter linewidths, while no changes were observed in decay rate. 
%Thus, while in the past HBT experiments alone have been presented as evidence for superradiance~\cite{QD_SR_1_2018, QD_SR_2_2019, SR_SV_1_2016}, 
Therefore, our experiments reveal that correlations are still present at very large detunings, where no change in the decay rate is observed, or expected.

D.~H, M.S.S and L.R.W acknowledge support from UK EPSRC (Grant No. EP/V026496/1). M.~C. is supported by the Return Program of the State of North Rhine-Westphalia. E.M.G. acknowledges support from UK EPSRC (Grant No. EP/T01377X/1) and from the Leverhulme Trust (RPG-2022-335). \\ \\
%%
%\textbf{Author contributions:} 
%\section*{Author contributions}
D.~H. and L.~H performed the experiments.
J.~W., M.~C. and E.M.G. developed the theoretical model.
D.~H, L.R.W, S.~S and A.~F contributed to the conception of the project.
L.~H., D.~H. and N.~M. designed the photonic devices.
R.~D. fabricated the photonic devices. 
I.~F. and A.K.V. carried out the growth of the wafer. 
D.~H. and J.~W. wrote the manuscript, with contributions from all co-authors.
M.S.S. and L.R.W supervised the project.
%
%\subsection{Photon counting measurements}
%Time dependent measurements are taken using two superconducting nanowire single-photon detectors. These have a combined time resolution of 30ps.\\

\bibliography{Bib_Main}

\begin{thebibliography}{41}%
\makeatletter
\providecommand \@ifxundefined [1]{%
 \@ifx{#1\undefined}
}%
\providecommand \@ifnum [1]{%
 \ifnum #1\expandafter \@firstoftwo
 \else \expandafter \@secondoftwo
 \fi
}%
\providecommand \@ifx [1]{%
 \ifx #1\expandafter \@firstoftwo
 \else \expandafter \@secondoftwo
 \fi
}%
\providecommand \natexlab [1]{#1}%
\providecommand \enquote  [1]{``#1''}%
\providecommand \bibnamefont  [1]{#1}%
\providecommand \bibfnamefont [1]{#1}%
\providecommand \citenamefont [1]{#1}%
\providecommand \href@noop [0]{\@secondoftwo}%
\providecommand \href [0]{\begingroup \@sanitize@url \@href}%
\providecommand \@href[1]{\@@startlink{#1}\@@href}%
\providecommand \@@href[1]{\endgroup#1\@@endlink}%
\providecommand \@sanitize@url [0]{\catcode `\\12\catcode `\$12\catcode
  `\&12\catcode `\#12\catcode `\^12\catcode `\_12\catcode `\%12\relax}%
\providecommand \@@startlink[1]{}%
\providecommand \@@endlink[0]{}%
\providecommand \url  [0]{\begingroup\@sanitize@url \@url }%
\providecommand \@url [1]{\endgroup\@href {#1}{\urlprefix }}%
\providecommand \urlprefix  [0]{URL }%
\providecommand \Eprint [0]{\href }%
\providecommand \doibase [0]{https://doi.org/}%
\providecommand \selectlanguage [0]{\@gobble}%
\providecommand \bibinfo  [0]{\@secondoftwo}%
\providecommand \bibfield  [0]{\@secondoftwo}%
\providecommand \translation [1]{[#1]}%
\providecommand \BibitemOpen [0]{}%
\providecommand \bibitemStop [0]{}%
\providecommand \bibitemNoStop [0]{.\EOS\space}%
\providecommand \EOS [0]{\spacefactor3000\relax}%
\providecommand \BibitemShut  [1]{\csname bibitem#1\endcsname}%
\let\auto@bib@innerbib\@empty
%</preamble>
\bibitem [{\citenamefont {Dicke}(1954{\natexlab{a}})}]{Dicke1954}%
  \BibitemOpen
  \bibfield  {author} {\bibinfo {author} {\bibfnamefont {R.~H.}\ \bibnamefont
  {Dicke}},\ }\bibfield  {title} {\bibinfo {title} {Coherence in spontaneous
  radiation processes},\ }\href {https://doi.org/10.1103/PhysRev.93.99}
  {\bibfield  {journal} {\bibinfo  {journal} {Phys. Rev.}\ }\textbf {\bibinfo
  {volume} {93}},\ \bibinfo {pages} {99} (\bibinfo {year}
  {1954}{\natexlab{a}})}\BibitemShut {NoStop}%
\bibitem [{\citenamefont {Goban}\ \emph {et~al.}(2015)\citenamefont {Goban},
  \citenamefont {Hung}, \citenamefont {Hood}, \citenamefont {Yu}, \citenamefont
  {Muniz}, \citenamefont {Painter},\ and\ \citenamefont
  {Kimble}}]{PhysRevLett.115.063601}%
  \BibitemOpen
  \bibfield  {author} {\bibinfo {author} {\bibfnamefont {A.}~\bibnamefont
  {Goban}}, \bibinfo {author} {\bibfnamefont {C.-L.}\ \bibnamefont {Hung}},
  \bibinfo {author} {\bibfnamefont {J.~D.}\ \bibnamefont {Hood}}, \bibinfo
  {author} {\bibfnamefont {S.-P.}\ \bibnamefont {Yu}}, \bibinfo {author}
  {\bibfnamefont {J.~A.}\ \bibnamefont {Muniz}}, \bibinfo {author}
  {\bibfnamefont {O.}~\bibnamefont {Painter}},\ and\ \bibinfo {author}
  {\bibfnamefont {H.~J.}\ \bibnamefont {Kimble}},\ }\bibfield  {title}
  {\bibinfo {title} {Superradiance for atoms trapped along a photonic crystal
  waveguide},\ }\href {https://doi.org/10.1103/PhysRevLett.115.063601}
  {\bibfield  {journal} {\bibinfo  {journal} {Phys. Rev. Lett.}\ }\textbf
  {\bibinfo {volume} {115}},\ \bibinfo {pages} {063601} (\bibinfo {year}
  {2015})}\BibitemShut {NoStop}%
\bibitem [{\citenamefont {Temnov}\ and\ \citenamefont
  {Woggon}(2009)}]{Atom_SR_Theory_1}%
  \BibitemOpen
  \bibfield  {author} {\bibinfo {author} {\bibfnamefont {V.~V.}\ \bibnamefont
  {Temnov}}\ and\ \bibinfo {author} {\bibfnamefont {U.}~\bibnamefont
  {Woggon}},\ }\bibfield  {title} {\bibinfo {title} {Photon statistics in the
  cooperative spontaneous emission},\ }\href
  {https://doi.org/10.1364/OE.17.005774} {\bibfield  {journal} {\bibinfo
  {journal} {Opt. Express}\ }\textbf {\bibinfo {volume} {17}},\ \bibinfo
  {pages} {5774} (\bibinfo {year} {2009})}\BibitemShut {NoStop}%
\bibitem [{\citenamefont {Imamo{\u g}lu}\ \emph {et~al.}(1999)\citenamefont
  {Imamo{\u g}lu}, \citenamefont {Awschalom}, \citenamefont {Burkard},
  \citenamefont {DiVincenzo}, \citenamefont {Loss}, \citenamefont {Sherwin},\
  and\ \citenamefont {Small}}]{PhysRevLett.83.4204}%
  \BibitemOpen
  \bibfield  {author} {\bibinfo {author} {\bibfnamefont {A.}~\bibnamefont
  {Imamo{\u g}lu}}, \bibinfo {author} {\bibfnamefont {D.~D.}\ \bibnamefont
  {Awschalom}}, \bibinfo {author} {\bibfnamefont {G.}~\bibnamefont {Burkard}},
  \bibinfo {author} {\bibfnamefont {D.~P.}\ \bibnamefont {DiVincenzo}},
  \bibinfo {author} {\bibfnamefont {D.}~\bibnamefont {Loss}}, \bibinfo {author}
  {\bibfnamefont {M.}~\bibnamefont {Sherwin}},\ and\ \bibinfo {author}
  {\bibfnamefont {A.}~\bibnamefont {Small}},\ }\bibfield  {title} {\bibinfo
  {title} {Quantum information processing using quantum dot spins and cavity
  qed},\ }\href {https://doi.org/10.1103/PhysRevLett.83.4204} {\bibfield
  {journal} {\bibinfo  {journal} {Phys. Rev. Lett.}\ }\textbf {\bibinfo
  {volume} {83}},\ \bibinfo {pages} {4204} (\bibinfo {year}
  {1999})}\BibitemShut {NoStop}%
\bibitem [{\citenamefont {Aspuru-Guzik}\ and\ \citenamefont
  {Walther}(2012)}]{Aspuru-Guzik2012}%
  \BibitemOpen
  \bibfield  {author} {\bibinfo {author} {\bibfnamefont {A.}~\bibnamefont
  {Aspuru-Guzik}}\ and\ \bibinfo {author} {\bibfnamefont {P.}~\bibnamefont
  {Walther}},\ }\bibfield  {title} {\bibinfo {title} {Photonic quantum
  simulators},\ }\href {https://doi.org/10.1038/nphys2253} {\bibfield
  {journal} {\bibinfo  {journal} {Nature Physics}\ }\textbf {\bibinfo {volume}
  {8}},\ \bibinfo {pages} {285} (\bibinfo {year} {2012})}\BibitemShut {NoStop}%
\bibitem [{\citenamefont {Liu}\ and\ \citenamefont
  {Tombesi}(1991)}]{W_S_Liu_1991}%
  \BibitemOpen
  \bibfield  {author} {\bibinfo {author} {\bibfnamefont {W.~S.}\ \bibnamefont
  {Liu}}\ and\ \bibinfo {author} {\bibfnamefont {P.}~\bibnamefont {Tombesi}},\
  }\bibfield  {title} {\bibinfo {title} {Squeezing in a super-radiant state},\
  }\href {https://doi.org/10.1088/0954-8998/3/2/002} {\bibfield  {journal}
  {\bibinfo  {journal} {Quantum Optics: Journal of the European Optical Society
  Part B}\ }\textbf {\bibinfo {volume} {3}},\ \bibinfo {pages} {93} (\bibinfo
  {year} {1991})}\BibitemShut {NoStop}%
\bibitem [{\citenamefont {Pagel}\ \emph {et~al.}(2015)\citenamefont {Pagel},
  \citenamefont {Alvermann},\ and\ \citenamefont
  {Fehske}}]{PhysRevA.91.043814}%
  \BibitemOpen
  \bibfield  {author} {\bibinfo {author} {\bibfnamefont {D.}~\bibnamefont
  {Pagel}}, \bibinfo {author} {\bibfnamefont {A.}~\bibnamefont {Alvermann}},\
  and\ \bibinfo {author} {\bibfnamefont {H.}~\bibnamefont {Fehske}},\
  }\bibfield  {title} {\bibinfo {title} {Nonclassical light from few emitters
  in a cavity},\ }\href {https://doi.org/10.1103/PhysRevA.91.043814} {\bibfield
   {journal} {\bibinfo  {journal} {Phys. Rev. A}\ }\textbf {\bibinfo {volume}
  {91}},\ \bibinfo {pages} {043814} (\bibinfo {year} {2015})}\BibitemShut
  {NoStop}%
\bibitem [{\citenamefont {Mlynek}\ \emph {et~al.}(2014)\citenamefont {Mlynek},
  \citenamefont {Abdumalikov}, \citenamefont {Eichler},\ and\ \citenamefont
  {Wallraff}}]{Mlynek2014}%
  \BibitemOpen
  \bibfield  {author} {\bibinfo {author} {\bibfnamefont {J.~A.}\ \bibnamefont
  {Mlynek}}, \bibinfo {author} {\bibfnamefont {A.~A.}\ \bibnamefont
  {Abdumalikov}}, \bibinfo {author} {\bibfnamefont {C.}~\bibnamefont
  {Eichler}},\ and\ \bibinfo {author} {\bibfnamefont {A.}~\bibnamefont
  {Wallraff}},\ }\bibfield  {title} {\bibinfo {title} {Observation of dicke
  superradiance for two artificial atoms in a cavity with high decay rate},\
  }\href {https://doi.org/10.1038/ncomms6186} {\bibfield  {journal} {\bibinfo
  {journal} {Nature Communications}\ }\textbf {\bibinfo {volume} {5}},\
  \bibinfo {pages} {5186} (\bibinfo {year} {2014})}\BibitemShut {NoStop}%
\bibitem [{\citenamefont {Gonz\'alez-Tudela}\ and\ \citenamefont
  {Porras}(2013)}]{PhysRevLett.110.080502}%
  \BibitemOpen
  \bibfield  {author} {\bibinfo {author} {\bibfnamefont {A.}~\bibnamefont
  {Gonz\'alez-Tudela}}\ and\ \bibinfo {author} {\bibfnamefont {D.}~\bibnamefont
  {Porras}},\ }\bibfield  {title} {\bibinfo {title} {Mesoscopic entanglement
  induced by spontaneous emission in solid-state quantum optics},\ }\href
  {https://doi.org/10.1103/PhysRevLett.110.080502} {\bibfield  {journal}
  {\bibinfo  {journal} {Phys. Rev. Lett.}\ }\textbf {\bibinfo {volume} {110}},\
  \bibinfo {pages} {080502} (\bibinfo {year} {2013})}\BibitemShut {NoStop}%
\bibitem [{\citenamefont {Julsgaard}\ and\ \citenamefont
  {M\o{}lmer}(2012)}]{PhysRevA.85.032327}%
  \BibitemOpen
  \bibfield  {author} {\bibinfo {author} {\bibfnamefont {B.}~\bibnamefont
  {Julsgaard}}\ and\ \bibinfo {author} {\bibfnamefont {K.}~\bibnamefont
  {M\o{}lmer}},\ }\bibfield  {title} {\bibinfo {title} {Measurement-induced
  two-qubit entanglement in a bad cavity: Fundamental and practical
  considerations},\ }\href {https://doi.org/10.1103/PhysRevA.85.032327}
  {\bibfield  {journal} {\bibinfo  {journal} {Phys. Rev. A}\ }\textbf {\bibinfo
  {volume} {85}},\ \bibinfo {pages} {032327} (\bibinfo {year}
  {2012})}\BibitemShut {NoStop}%
\bibitem [{\citenamefont {Gonzalez-Ballestero}\ \emph
  {et~al.}(2015)\citenamefont {Gonzalez-Ballestero}, \citenamefont
  {Gonzalez-Tudela}, \citenamefont {Garcia-Vidal},\ and\ \citenamefont
  {Moreno}}]{Chiral_SR_1_2015}%
  \BibitemOpen
  \bibfield  {author} {\bibinfo {author} {\bibfnamefont {C.}~\bibnamefont
  {Gonzalez-Ballestero}}, \bibinfo {author} {\bibfnamefont {A.}~\bibnamefont
  {Gonzalez-Tudela}}, \bibinfo {author} {\bibfnamefont {F.~J.}\ \bibnamefont
  {Garcia-Vidal}},\ and\ \bibinfo {author} {\bibfnamefont {E.}~\bibnamefont
  {Moreno}},\ }\bibfield  {title} {\bibinfo {title} {Chiral route to
  spontaneous entanglement generation},\ }\href
  {https://doi.org/10.1103/PhysRevB.92.155304} {\bibfield  {journal} {\bibinfo
  {journal} {Phys. Rev. B}\ }\textbf {\bibinfo {volume} {92}},\ \bibinfo
  {pages} {155304} (\bibinfo {year} {2015})}\BibitemShut {NoStop}%
\bibitem [{\citenamefont {Buonaiuto}\ \emph {et~al.}(2019)\citenamefont
  {Buonaiuto}, \citenamefont {Jones}, \citenamefont {Olmos},\ and\
  \citenamefont {Lesanovsky}}]{Chiral_SR_4_2019}%
  \BibitemOpen
  \bibfield  {author} {\bibinfo {author} {\bibfnamefont {G.}~\bibnamefont
  {Buonaiuto}}, \bibinfo {author} {\bibfnamefont {R.}~\bibnamefont {Jones}},
  \bibinfo {author} {\bibfnamefont {B.}~\bibnamefont {Olmos}},\ and\ \bibinfo
  {author} {\bibfnamefont {I.}~\bibnamefont {Lesanovsky}},\ }\bibfield  {title}
  {\bibinfo {title} {Dynamical creation and detection of entangled many-body
  states in a chiral atom chain},\ }\href
  {https://doi.org/10.1088/1367-2630/ab4f50} {\bibfield  {journal} {\bibinfo
  {journal} {New Journal of Physics}\ }\textbf {\bibinfo {volume} {21}},\
  \bibinfo {pages} {113021} (\bibinfo {year} {2019})}\BibitemShut {NoStop}%
\bibitem [{\citenamefont {Politi}\ \emph {et~al.}(2009)\citenamefont {Politi},
  \citenamefont {Matthews},\ and\ \citenamefont
  {O'Brien}}]{doi:10.1126/science.1173731}%
  \BibitemOpen
  \bibfield  {author} {\bibinfo {author} {\bibfnamefont {A.}~\bibnamefont
  {Politi}}, \bibinfo {author} {\bibfnamefont {J.~C.~F.}\ \bibnamefont
  {Matthews}},\ and\ \bibinfo {author} {\bibfnamefont {J.~L.}\ \bibnamefont
  {O'Brien}},\ }\bibfield  {title} {\bibinfo {title} {Shor’s quantum
  factoring algorithm on a photonic chip},\ }\href
  {https://doi.org/10.1126/science.1173731} {\bibfield  {journal} {\bibinfo
  {journal} {Science}\ }\textbf {\bibinfo {volume} {325}},\ \bibinfo {pages}
  {1221} (\bibinfo {year} {2009})}\BibitemShut {NoStop}%
\bibitem [{\citenamefont {Chang}\ \emph {et~al.}(2018)\citenamefont {Chang},
  \citenamefont {Douglas}, \citenamefont {Gonz\'alez-Tudela}, \citenamefont
  {Hung},\ and\ \citenamefont {Kimble}}]{RevModPhys.90.031002}%
  \BibitemOpen
  \bibfield  {author} {\bibinfo {author} {\bibfnamefont {D.~E.}\ \bibnamefont
  {Chang}}, \bibinfo {author} {\bibfnamefont {J.~S.}\ \bibnamefont {Douglas}},
  \bibinfo {author} {\bibfnamefont {A.}~\bibnamefont {Gonz\'alez-Tudela}},
  \bibinfo {author} {\bibfnamefont {C.-L.}\ \bibnamefont {Hung}},\ and\
  \bibinfo {author} {\bibfnamefont {H.~J.}\ \bibnamefont {Kimble}},\ }\bibfield
   {title} {\bibinfo {title} {Colloquium: Quantum matter built from nanoscopic
  lattices of atoms and photons},\ }\href
  {https://doi.org/10.1103/RevModPhys.90.031002} {\bibfield  {journal}
  {\bibinfo  {journal} {Rev. Mod. Phys.}\ }\textbf {\bibinfo {volume} {90}},\
  \bibinfo {pages} {031002} (\bibinfo {year} {2018})}\BibitemShut {NoStop}%
\bibitem [{\citenamefont {Quach}\ \emph {et~al.}(2022)\citenamefont {Quach},
  \citenamefont {McGhee}, \citenamefont {Ganzer}, \citenamefont {Rouse},
  \citenamefont {Lovett}, \citenamefont {Gauger}, \citenamefont {Keeling},
  \citenamefont {Cerullo}, \citenamefont {Lidzey},\ and\ \citenamefont
  {Virgili}}]{quach_superabsorption_2022}%
  \BibitemOpen
  \bibfield  {author} {\bibinfo {author} {\bibfnamefont {J.~Q.}\ \bibnamefont
  {Quach}}, \bibinfo {author} {\bibfnamefont {K.~E.}\ \bibnamefont {McGhee}},
  \bibinfo {author} {\bibfnamefont {L.}~\bibnamefont {Ganzer}}, \bibinfo
  {author} {\bibfnamefont {D.~M.}\ \bibnamefont {Rouse}}, \bibinfo {author}
  {\bibfnamefont {B.~W.}\ \bibnamefont {Lovett}}, \bibinfo {author}
  {\bibfnamefont {E.~M.}\ \bibnamefont {Gauger}}, \bibinfo {author}
  {\bibfnamefont {J.}~\bibnamefont {Keeling}}, \bibinfo {author} {\bibfnamefont
  {G.}~\bibnamefont {Cerullo}}, \bibinfo {author} {\bibfnamefont {D.~G.}\
  \bibnamefont {Lidzey}},\ and\ \bibinfo {author} {\bibfnamefont
  {T.}~\bibnamefont {Virgili}},\ }\bibfield  {title} {\bibinfo {title}
  {Superabsorption in an organic microcavity: Toward a quantum battery},\
  }\href {https://doi.org/10.1126/sciadv.abk3160} {\bibfield  {journal}
  {\bibinfo  {journal} {Science Advances}\ }\textbf {\bibinfo {volume} {8}},\
  \bibinfo {pages} {eabk3160} (\bibinfo {year} {2022})}\BibitemShut {NoStop}%
\bibitem [{\citenamefont {Rastogi}\ \emph {et~al.}(2022)\citenamefont
  {Rastogi}, \citenamefont {Saglamyurek}, \citenamefont {Hrushevskyi},\ and\
  \citenamefont {LeBlanc}}]{PhysRevLett.129.120502}%
  \BibitemOpen
  \bibfield  {author} {\bibinfo {author} {\bibfnamefont {A.}~\bibnamefont
  {Rastogi}}, \bibinfo {author} {\bibfnamefont {E.}~\bibnamefont
  {Saglamyurek}}, \bibinfo {author} {\bibfnamefont {T.}~\bibnamefont
  {Hrushevskyi}},\ and\ \bibinfo {author} {\bibfnamefont {L.~J.}\ \bibnamefont
  {LeBlanc}},\ }\bibfield  {title} {\bibinfo {title} {Superradiance-mediated
  photon storage for broadband quantum memory},\ }\href
  {https://doi.org/10.1103/PhysRevLett.129.120502} {\bibfield  {journal}
  {\bibinfo  {journal} {Phys. Rev. Lett.}\ }\textbf {\bibinfo {volume} {129}},\
  \bibinfo {pages} {120502} (\bibinfo {year} {2022})}\BibitemShut {NoStop}%
\bibitem [{\citenamefont {Kalachev}(2008)}]{Application_QMem}%
  \BibitemOpen
  \bibfield  {author} {\bibinfo {author} {\bibfnamefont {A.~A.}\ \bibnamefont
  {Kalachev}},\ }\bibfield  {title} {\bibinfo {title} {Quantum memory based on
  optical subradiance: Optimization of the signal-to-noise ratio},\ }\href
  {https://doi.org/10.3103/S1062873808050286} {\bibfield  {journal} {\bibinfo
  {journal} {Bulletin of the Russian Academy of Sciences: Physics}\ }\textbf
  {\bibinfo {volume} {72}},\ \bibinfo {pages} {691} (\bibinfo {year}
  {2008})}\BibitemShut {NoStop}%
\bibitem [{\citenamefont {Davidson}\ \emph {et~al.}(2022)\citenamefont
  {Davidson}, \citenamefont {Pollock},\ and\ \citenamefont
  {Gauger}}]{PRXQuantum.3.020354}%
  \BibitemOpen
  \bibfield  {author} {\bibinfo {author} {\bibfnamefont {S.}~\bibnamefont
  {Davidson}}, \bibinfo {author} {\bibfnamefont {F.~A.}\ \bibnamefont
  {Pollock}},\ and\ \bibinfo {author} {\bibfnamefont {E.}~\bibnamefont
  {Gauger}},\ }\bibfield  {title} {\bibinfo {title} {Eliminating radiative
  losses in long-range exciton transport},\ }\href
  {https://doi.org/10.1103/PRXQuantum.3.020354} {\bibfield  {journal} {\bibinfo
   {journal} {PRX Quantum}\ }\textbf {\bibinfo {volume} {3}},\ \bibinfo {pages}
  {020354} (\bibinfo {year} {2022})}\BibitemShut {NoStop}%
\bibitem [{\citenamefont {Mattioni}\ \emph {et~al.}(2021)\citenamefont
  {Mattioni}, \citenamefont {Caycedo-Soler}, \citenamefont {Huelga},\ and\
  \citenamefont {Plenio}}]{PhysRevX.11.041003}%
  \BibitemOpen
  \bibfield  {author} {\bibinfo {author} {\bibfnamefont {A.}~\bibnamefont
  {Mattioni}}, \bibinfo {author} {\bibfnamefont {F.}~\bibnamefont
  {Caycedo-Soler}}, \bibinfo {author} {\bibfnamefont {S.~F.}\ \bibnamefont
  {Huelga}},\ and\ \bibinfo {author} {\bibfnamefont {M.~B.}\ \bibnamefont
  {Plenio}},\ }\bibfield  {title} {\bibinfo {title} {Design principles for
  long-range energy transfer at room temperature},\ }\href
  {https://doi.org/10.1103/PhysRevX.11.041003} {\bibfield  {journal} {\bibinfo
  {journal} {Phys. Rev. X}\ }\textbf {\bibinfo {volume} {11}},\ \bibinfo
  {pages} {041003} (\bibinfo {year} {2021})}\BibitemShut {NoStop}%
\bibitem [{\citenamefont {Dicke}(1954{\natexlab{b}})}]{Dicke1}%
  \BibitemOpen
  \bibfield  {author} {\bibinfo {author} {\bibfnamefont {R.~H.}\ \bibnamefont
  {Dicke}},\ }\bibfield  {title} {\bibinfo {title} {Coherence in spontaneous
  radiation processes},\ }\href {https://doi.org/10.1103/PhysRev.93.99}
  {\bibfield  {journal} {\bibinfo  {journal} {Phys. Rev.}\ }\textbf {\bibinfo
  {volume} {93}},\ \bibinfo {pages} {99} (\bibinfo {year}
  {1954}{\natexlab{b}})}\BibitemShut {NoStop}%
\bibitem [{\citenamefont {Gross}\ and\ \citenamefont
  {Haroche}(1982)}]{gross_superradiance_1982}%
  \BibitemOpen
  \bibfield  {author} {\bibinfo {author} {\bibfnamefont {M.}~\bibnamefont
  {Gross}}\ and\ \bibinfo {author} {\bibfnamefont {S.}~\bibnamefont
  {Haroche}},\ }\bibfield  {title} {\bibinfo {title} {Superradiance: {An} essay
  on the theory of collective spontaneous emission},\ }\href
  {https://doi.org/10.1016/0370-1573(82)90102-8} {\bibfield  {journal}
  {\bibinfo  {journal} {Physics Reports}\ }\textbf {\bibinfo {volume} {93}},\
  \bibinfo {pages} {301} (\bibinfo {year} {1982})}\BibitemShut {NoStop}%
\bibitem [{\citenamefont {Tiranov}\ \emph {et~al.}(2023)\citenamefont
  {Tiranov}, \citenamefont {Angelopoulou}, \citenamefont {van Diepen},
  \citenamefont {Schrinski}, \citenamefont {Sandberg}, \citenamefont {Wang},
  \citenamefont {Midolo}, \citenamefont {Scholz}, \citenamefont {Wieck},
  \citenamefont {Ludwig}, \citenamefont {Sørensen},\ and\ \citenamefont
  {Lodahl}}]{QD_DR_3_2023}%
  \BibitemOpen
  \bibfield  {author} {\bibinfo {author} {\bibfnamefont {A.}~\bibnamefont
  {Tiranov}}, \bibinfo {author} {\bibfnamefont {V.}~\bibnamefont
  {Angelopoulou}}, \bibinfo {author} {\bibfnamefont {C.~J.}\ \bibnamefont {van
  Diepen}}, \bibinfo {author} {\bibfnamefont {B.}~\bibnamefont {Schrinski}},
  \bibinfo {author} {\bibfnamefont {O.~A.~D.}\ \bibnamefont {Sandberg}},
  \bibinfo {author} {\bibfnamefont {Y.}~\bibnamefont {Wang}}, \bibinfo {author}
  {\bibfnamefont {L.}~\bibnamefont {Midolo}}, \bibinfo {author} {\bibfnamefont
  {S.}~\bibnamefont {Scholz}}, \bibinfo {author} {\bibfnamefont {A.~D.}\
  \bibnamefont {Wieck}}, \bibinfo {author} {\bibfnamefont {A.}~\bibnamefont
  {Ludwig}}, \bibinfo {author} {\bibfnamefont {A.~S.}\ \bibnamefont
  {Sørensen}},\ and\ \bibinfo {author} {\bibfnamefont {P.}~\bibnamefont
  {Lodahl}},\ }\bibfield  {title} {\bibinfo {title} {Collective super- and
  subradiant dynamics between distant optical quantum emitters},\ }\href
  {https://doi.org/10.1126/science.ade9324} {\bibfield  {journal} {\bibinfo
  {journal} {Science}\ }\textbf {\bibinfo {volume} {379}},\ \bibinfo {pages}
  {389} (\bibinfo {year} {2023})}\BibitemShut {NoStop}%
\bibitem [{\citenamefont {Wiercinski}\ \emph {et~al.}(2024)\citenamefont
  {Wiercinski}, \citenamefont {Cygorek},\ and\ \citenamefont
  {Gauger}}]{PhysRevResearch.6.033231}%
  \BibitemOpen
  \bibfield  {author} {\bibinfo {author} {\bibfnamefont {J.}~\bibnamefont
  {Wiercinski}}, \bibinfo {author} {\bibfnamefont {M.}~\bibnamefont
  {Cygorek}},\ and\ \bibinfo {author} {\bibfnamefont {E.~M.}\ \bibnamefont
  {Gauger}},\ }\bibfield  {title} {\bibinfo {title} {Role of polaron dressing
  in superradiant emission dynamics},\ }\href
  {https://doi.org/10.1103/PhysRevResearch.6.033231} {\bibfield  {journal}
  {\bibinfo  {journal} {Phys. Rev. Res.}\ }\textbf {\bibinfo {volume} {6}},\
  \bibinfo {pages} {033231} (\bibinfo {year} {2024})}\BibitemShut {NoStop}%
\bibitem [{\citenamefont {Scheibner}\ \emph {et~al.}(2007)\citenamefont
  {Scheibner}, \citenamefont {Schmidt}, \citenamefont {Worschech},
  \citenamefont {Forchel}, \citenamefont {Bacher}, \citenamefont {Passow},\
  and\ \citenamefont {Hommel}}]{scheibner_superradiance_2007}%
  \BibitemOpen
  \bibfield  {author} {\bibinfo {author} {\bibfnamefont {M.}~\bibnamefont
  {Scheibner}}, \bibinfo {author} {\bibfnamefont {T.}~\bibnamefont {Schmidt}},
  \bibinfo {author} {\bibfnamefont {L.}~\bibnamefont {Worschech}}, \bibinfo
  {author} {\bibfnamefont {A.}~\bibnamefont {Forchel}}, \bibinfo {author}
  {\bibfnamefont {G.}~\bibnamefont {Bacher}}, \bibinfo {author} {\bibfnamefont
  {T.}~\bibnamefont {Passow}},\ and\ \bibinfo {author} {\bibfnamefont
  {D.}~\bibnamefont {Hommel}},\ }\bibfield  {title} {\bibinfo {title}
  {Superradiance of quantum dots},\ }\href {https://doi.org/10.1038/nphys494}
  {\bibfield  {journal} {\bibinfo  {journal} {Nature Physics}\ }\textbf
  {\bibinfo {volume} {3}},\ \bibinfo {pages} {106} (\bibinfo {year}
  {2007})}\BibitemShut {NoStop}%
\bibitem [{\citenamefont {Bhatti}\ \emph {et~al.}(2015)\citenamefont {Bhatti},
  \citenamefont {von Zanthier},\ and\ \citenamefont
  {Agarwal}}]{bhatti_superbunching_2015}%
  \BibitemOpen
  \bibfield  {author} {\bibinfo {author} {\bibfnamefont {D.}~\bibnamefont
  {Bhatti}}, \bibinfo {author} {\bibfnamefont {J.}~\bibnamefont {von
  Zanthier}},\ and\ \bibinfo {author} {\bibfnamefont {G.~S.}\ \bibnamefont
  {Agarwal}},\ }\bibfield  {title} {\bibinfo {title} {Superbunching and
  {Nonclassicality} as new {Hallmarks} of {Superradiance}},\ }\href
  {https://doi.org/10.1038/srep17335} {\bibfield  {journal} {\bibinfo
  {journal} {Scientific Reports}\ }\textbf {\bibinfo {volume} {5}},\ \bibinfo
  {pages} {17335} (\bibinfo {year} {2015})}\BibitemShut {NoStop}%
\bibitem [{\citenamefont {Cygorek}\ \emph {et~al.}(2023)\citenamefont
  {Cygorek}, \citenamefont {Scerri}, \citenamefont {Santana}, \citenamefont
  {Koong}, \citenamefont {Gerardot},\ and\ \citenamefont
  {Gauger}}]{Coop_Theory1_2023}%
  \BibitemOpen
  \bibfield  {author} {\bibinfo {author} {\bibfnamefont {M.}~\bibnamefont
  {Cygorek}}, \bibinfo {author} {\bibfnamefont {E.~D.}\ \bibnamefont {Scerri}},
  \bibinfo {author} {\bibfnamefont {T.~S.}\ \bibnamefont {Santana}}, \bibinfo
  {author} {\bibfnamefont {Z.~X.}\ \bibnamefont {Koong}}, \bibinfo {author}
  {\bibfnamefont {B.~D.}\ \bibnamefont {Gerardot}},\ and\ \bibinfo {author}
  {\bibfnamefont {E.~M.}\ \bibnamefont {Gauger}},\ }\bibfield  {title}
  {\bibinfo {title} {Signatures of cooperative emission in photon coincidence:
  Superradiance versus measurement-induced cooperativity},\ }\href
  {https://doi.org/10.1103/PhysRevA.107.023718} {\bibfield  {journal} {\bibinfo
   {journal} {Phys. Rev. A}\ }\textbf {\bibinfo {volume} {107}},\ \bibinfo
  {pages} {023718} (\bibinfo {year} {2023})}\BibitemShut {NoStop}%
\bibitem [{\citenamefont {Kim}\ \emph {et~al.}(2016)\citenamefont {Kim},
  \citenamefont {Richardson}, \citenamefont {Leavitt},\ and\ \citenamefont
  {Waks}}]{doi:10.1021/acs.nanolett.6b03295}%
  \BibitemOpen
  \bibfield  {author} {\bibinfo {author} {\bibfnamefont {J.-H.}\ \bibnamefont
  {Kim}}, \bibinfo {author} {\bibfnamefont {C.~J.~K.}\ \bibnamefont
  {Richardson}}, \bibinfo {author} {\bibfnamefont {R.~P.}\ \bibnamefont
  {Leavitt}},\ and\ \bibinfo {author} {\bibfnamefont {E.}~\bibnamefont
  {Waks}},\ }\bibfield  {title} {\bibinfo {title} {Two-photon interference from
  the far-field emission of chip-integrated cavity-coupled emitters},\ }\href
  {https://doi.org/10.1021/acs.nanolett.6b03295} {\bibfield  {journal}
  {\bibinfo  {journal} {Nano Letters}\ }\textbf {\bibinfo {volume} {16}},\
  \bibinfo {pages} {7061} (\bibinfo {year} {2016})}\BibitemShut {NoStop}%
\bibitem [{\citenamefont {Wolf}\ \emph {et~al.}(2020)\citenamefont {Wolf},
  \citenamefont {Richter}, \citenamefont {von Zanthier},\ and\ \citenamefont
  {Schmidt-Kaler}}]{PhysRevLett.124.063603}%
  \BibitemOpen
  \bibfield  {author} {\bibinfo {author} {\bibfnamefont {S.}~\bibnamefont
  {Wolf}}, \bibinfo {author} {\bibfnamefont {S.}~\bibnamefont {Richter}},
  \bibinfo {author} {\bibfnamefont {J.}~\bibnamefont {von Zanthier}},\ and\
  \bibinfo {author} {\bibfnamefont {F.}~\bibnamefont {Schmidt-Kaler}},\
  }\bibfield  {title} {\bibinfo {title} {Light of two atoms in free space:
  Bunching or antibunching?},\ }\href
  {https://doi.org/10.1103/PhysRevLett.124.063603} {\bibfield  {journal}
  {\bibinfo  {journal} {Phys. Rev. Lett.}\ }\textbf {\bibinfo {volume} {124}},\
  \bibinfo {pages} {063603} (\bibinfo {year} {2020})}\BibitemShut {NoStop}%
\bibitem [{\citenamefont {Koong}\ \emph {et~al.}(2022)\citenamefont {Koong},
  \citenamefont {Cygorek}, \citenamefont {Scerri}, \citenamefont {Santana},
  \citenamefont {Park}, \citenamefont {Song}, \citenamefont {Gauger},\ and\
  \citenamefont {Gerardot}}]{QD_SR_3_2022}%
  \BibitemOpen
  \bibfield  {author} {\bibinfo {author} {\bibfnamefont {Z.~X.}\ \bibnamefont
  {Koong}}, \bibinfo {author} {\bibfnamefont {M.}~\bibnamefont {Cygorek}},
  \bibinfo {author} {\bibfnamefont {E.}~\bibnamefont {Scerri}}, \bibinfo
  {author} {\bibfnamefont {T.~S.}\ \bibnamefont {Santana}}, \bibinfo {author}
  {\bibfnamefont {S.~I.}\ \bibnamefont {Park}}, \bibinfo {author}
  {\bibfnamefont {J.~D.}\ \bibnamefont {Song}}, \bibinfo {author}
  {\bibfnamefont {E.~M.}\ \bibnamefont {Gauger}},\ and\ \bibinfo {author}
  {\bibfnamefont {B.~D.}\ \bibnamefont {Gerardot}},\ }\bibfield  {title}
  {\bibinfo {title} {Coherence in cooperative photon emission from
  indistinguishable quantum emitters},\ }\href
  {https://doi.org/10.1126/sciadv.abm8171} {\bibfield  {journal} {\bibinfo
  {journal} {Science Advances}\ }\textbf {\bibinfo {volume} {8}},\ \bibinfo
  {pages} {eabm8171} (\bibinfo {year} {2022})}\BibitemShut {NoStop}%
\bibitem [{\citenamefont {Wiercinski}\ \emph {et~al.}(2023)\citenamefont
  {Wiercinski}, \citenamefont {Gauger},\ and\ \citenamefont
  {Cygorek}}]{wiercinski_phonon_2023}%
  \BibitemOpen
  \bibfield  {author} {\bibinfo {author} {\bibfnamefont {J.}~\bibnamefont
  {Wiercinski}}, \bibinfo {author} {\bibfnamefont {E.~M.}\ \bibnamefont
  {Gauger}},\ and\ \bibinfo {author} {\bibfnamefont {M.}~\bibnamefont
  {Cygorek}},\ }\bibfield  {title} {\bibinfo {title} {Phonon coupling versus
  pure dephasing in the photon statistics of cooperative emitters},\ }\href
  {https://doi.org/10.1103/PhysRevResearch.5.013176} {\bibfield  {journal}
  {\bibinfo  {journal} {Physical Review Research}\ }\textbf {\bibinfo {volume}
  {5}},\ \bibinfo {pages} {013176} (\bibinfo {year} {2023})},\ \bibinfo {note}
  {publisher: American Physical Society}\BibitemShut {NoStop}%
\bibitem [{\citenamefont {Zhai}\ \emph {et~al.}(2020)\citenamefont {Zhai},
  \citenamefont {Löbl}, \citenamefont {Nguyen}, \citenamefont {Ritzmann},
  \citenamefont {Javadi}, \citenamefont {Spinnler}, \citenamefont {Wieck},
  \citenamefont {Ludwig},\ and\ \citenamefont {Warburton}}]{Zhai2020}%
  \BibitemOpen
  \bibfield  {author} {\bibinfo {author} {\bibfnamefont {L.}~\bibnamefont
  {Zhai}}, \bibinfo {author} {\bibfnamefont {M.~C.}\ \bibnamefont {Löbl}},
  \bibinfo {author} {\bibfnamefont {G.~N.}\ \bibnamefont {Nguyen}}, \bibinfo
  {author} {\bibfnamefont {J.}~\bibnamefont {Ritzmann}}, \bibinfo {author}
  {\bibfnamefont {A.}~\bibnamefont {Javadi}}, \bibinfo {author} {\bibfnamefont
  {C.}~\bibnamefont {Spinnler}}, \bibinfo {author} {\bibfnamefont {A.~D.}\
  \bibnamefont {Wieck}}, \bibinfo {author} {\bibfnamefont {A.}~\bibnamefont
  {Ludwig}},\ and\ \bibinfo {author} {\bibfnamefont {R.~J.}\ \bibnamefont
  {Warburton}},\ }\bibfield  {title} {\bibinfo {title} {Low-noise gaas quantum
  dots for quantum photonics},\ }\href
  {https://doi.org/10.1038/s41467-020-18625-z} {\bibfield  {journal} {\bibinfo
  {journal} {Nat. Comms.}\ }\textbf {\bibinfo {volume} {11}},\ \bibinfo {pages}
  {4745} (\bibinfo {year} {2020})}\BibitemShut {NoStop}%
\bibitem [{\citenamefont {Zhai}\ \emph {et~al.}(2022)\citenamefont {Zhai},
  \citenamefont {Nguyen}, \citenamefont {Spinnler}, \citenamefont {Ritzmann},
  \citenamefont {Löbl}, \citenamefont {Wieck}, \citenamefont {Ludwig},
  \citenamefont {Javadi},\ and\ \citenamefont {Warburton}}]{Zhai2022}%
  \BibitemOpen
  \bibfield  {author} {\bibinfo {author} {\bibfnamefont {L.}~\bibnamefont
  {Zhai}}, \bibinfo {author} {\bibfnamefont {G.~N.}\ \bibnamefont {Nguyen}},
  \bibinfo {author} {\bibfnamefont {C.}~\bibnamefont {Spinnler}}, \bibinfo
  {author} {\bibfnamefont {J.}~\bibnamefont {Ritzmann}}, \bibinfo {author}
  {\bibfnamefont {M.~C.}\ \bibnamefont {Löbl}}, \bibinfo {author}
  {\bibfnamefont {A.~D.}\ \bibnamefont {Wieck}}, \bibinfo {author}
  {\bibfnamefont {A.}~\bibnamefont {Ludwig}}, \bibinfo {author} {\bibfnamefont
  {A.}~\bibnamefont {Javadi}},\ and\ \bibinfo {author} {\bibfnamefont {R.~J.}\
  \bibnamefont {Warburton}},\ }\bibfield  {title} {\bibinfo {title} {Quantum
  interference of identical photons from remote gaas quantum dots},\ }\href
  {https://doi.org/10.1038/s41467-020-18625-z} {\bibfield  {journal} {\bibinfo
  {journal} {Nature Nanotechnology}\ }\textbf {\bibinfo {volume} {17}},\
  \bibinfo {pages} {829} (\bibinfo {year} {2022})}\BibitemShut {NoStop}%
\bibitem [{\citenamefont {Lodahl}\ \emph {et~al.}(2015)\citenamefont {Lodahl},
  \citenamefont {Mahmoodian},\ and\ \citenamefont
  {Stobbe}}]{lodahl_interfacing_2015}%
  \BibitemOpen
  \bibfield  {author} {\bibinfo {author} {\bibfnamefont {P.}~\bibnamefont
  {Lodahl}}, \bibinfo {author} {\bibfnamefont {S.}~\bibnamefont {Mahmoodian}},\
  and\ \bibinfo {author} {\bibfnamefont {S.}~\bibnamefont {Stobbe}},\
  }\bibfield  {title} {\bibinfo {title} {Interfacing single photons and single
  quantum dots with photonic nanostructures},\ }\href
  {https://doi.org/10.1103/RevModPhys.87.347} {\bibfield  {journal} {\bibinfo
  {journal} {Reviews of Modern Physics}\ }\textbf {\bibinfo {volume} {87}},\
  \bibinfo {pages} {347} (\bibinfo {year} {2015})}\BibitemShut {NoStop}%
\bibitem [{\citenamefont {Arcari}\ \emph {et~al.}(2014)\citenamefont {Arcari},
  \citenamefont {S\"ollner}, \citenamefont {Javadi}, \citenamefont
  {Lindskov~Hansen}, \citenamefont {Mahmoodian}, \citenamefont {Liu},
  \citenamefont {Thyrrestrup}, \citenamefont {Lee}, \citenamefont {Song},
  \citenamefont {Stobbe},\ and\ \citenamefont
  {Lodahl}}]{PhysRevLett.113.093603}%
  \BibitemOpen
  \bibfield  {author} {\bibinfo {author} {\bibfnamefont {M.}~\bibnamefont
  {Arcari}}, \bibinfo {author} {\bibfnamefont {I.}~\bibnamefont {S\"ollner}},
  \bibinfo {author} {\bibfnamefont {A.}~\bibnamefont {Javadi}}, \bibinfo
  {author} {\bibfnamefont {S.}~\bibnamefont {Lindskov~Hansen}}, \bibinfo
  {author} {\bibfnamefont {S.}~\bibnamefont {Mahmoodian}}, \bibinfo {author}
  {\bibfnamefont {J.}~\bibnamefont {Liu}}, \bibinfo {author} {\bibfnamefont
  {H.}~\bibnamefont {Thyrrestrup}}, \bibinfo {author} {\bibfnamefont {E.~H.}\
  \bibnamefont {Lee}}, \bibinfo {author} {\bibfnamefont {J.~D.}\ \bibnamefont
  {Song}}, \bibinfo {author} {\bibfnamefont {S.}~\bibnamefont {Stobbe}},\ and\
  \bibinfo {author} {\bibfnamefont {P.}~\bibnamefont {Lodahl}},\ }\bibfield
  {title} {\bibinfo {title} {Near-unity coupling efficiency of a quantum
  emitter to a photonic crystal waveguide},\ }\href
  {https://doi.org/10.1103/PhysRevLett.113.093603} {\bibfield  {journal}
  {\bibinfo  {journal} {Phys. Rev. Lett.}\ }\textbf {\bibinfo {volume} {113}},\
  \bibinfo {pages} {093603} (\bibinfo {year} {2014})}\BibitemShut {NoStop}%
\bibitem [{\citenamefont {Kim}\ \emph {et~al.}(2018)\citenamefont {Kim},
  \citenamefont {Aghaeimeibodi}, \citenamefont {Richardson}, \citenamefont
  {Leavitt},\ and\ \citenamefont {Waks}}]{QD_SR_1_2018}%
  \BibitemOpen
  \bibfield  {author} {\bibinfo {author} {\bibfnamefont {J.-H.}\ \bibnamefont
  {Kim}}, \bibinfo {author} {\bibfnamefont {S.}~\bibnamefont {Aghaeimeibodi}},
  \bibinfo {author} {\bibfnamefont {C.~J.~K.}\ \bibnamefont {Richardson}},
  \bibinfo {author} {\bibfnamefont {R.~P.}\ \bibnamefont {Leavitt}},\ and\
  \bibinfo {author} {\bibfnamefont {E.}~\bibnamefont {Waks}},\ }\bibfield
  {title} {\bibinfo {title} {Super-radiant emission from quantum dots in a
  nanophotonic waveguide},\ }\href
  {https://doi.org/10.1021/acs.nanolett.8b01133} {\bibfield  {journal}
  {\bibinfo  {journal} {Nano Letters}\ }\textbf {\bibinfo {volume} {18}},\
  \bibinfo {pages} {4734} (\bibinfo {year} {2018})}\BibitemShut {NoStop}%
\bibitem [{\citenamefont {Grim}\ \emph {et~al.}(2019)\citenamefont {Grim},
  \citenamefont {Bracker}, \citenamefont {Zalalutdinov}, \citenamefont
  {Carter}, \citenamefont {Kozen}, \citenamefont {Kim}, \citenamefont {Kim},
  \citenamefont {Mlack}, \citenamefont {Yakes}, \citenamefont {Lee},\ and\
  \citenamefont {Gammon}}]{QD_SR_2_2019}%
  \BibitemOpen
  \bibfield  {author} {\bibinfo {author} {\bibfnamefont {J.~Q.}\ \bibnamefont
  {Grim}}, \bibinfo {author} {\bibfnamefont {A.~S.}\ \bibnamefont {Bracker}},
  \bibinfo {author} {\bibfnamefont {M.}~\bibnamefont {Zalalutdinov}}, \bibinfo
  {author} {\bibfnamefont {S.~G.}\ \bibnamefont {Carter}}, \bibinfo {author}
  {\bibfnamefont {A.~C.}\ \bibnamefont {Kozen}}, \bibinfo {author}
  {\bibfnamefont {M.}~\bibnamefont {Kim}}, \bibinfo {author} {\bibfnamefont
  {C.~S.}\ \bibnamefont {Kim}}, \bibinfo {author} {\bibfnamefont {J.~T.}\
  \bibnamefont {Mlack}}, \bibinfo {author} {\bibfnamefont {M.}~\bibnamefont
  {Yakes}}, \bibinfo {author} {\bibfnamefont {B.}~\bibnamefont {Lee}},\ and\
  \bibinfo {author} {\bibfnamefont {D.}~\bibnamefont {Gammon}},\ }\bibfield
  {title} {\bibinfo {title} {Scalable in operando strain tuning in nanophotonic
  waveguides enabling three-quantum-dot superradiance},\ }\href
  {https://doi.org/10.1038/s41563-019-0418-0} {\bibfield  {journal} {\bibinfo
  {journal} {Nature Materials}\ }\textbf {\bibinfo {volume} {18}},\ \bibinfo
  {pages} {963} (\bibinfo {year} {2019})}\BibitemShut {NoStop}%
\bibitem [{\citenamefont {Sipahigil}\ \emph {et~al.}(2016)\citenamefont
  {Sipahigil}, \citenamefont {Evans}, \citenamefont {Sukachev}, \citenamefont
  {Burek}, \citenamefont {Borregaard}, \citenamefont {Bhaskar}, \citenamefont
  {Nguyen}, \citenamefont {Pacheco}, \citenamefont {Atikian}, \citenamefont
  {Meuwly}, \citenamefont {Camacho}, \citenamefont {Jelezko}, \citenamefont
  {Bielejec}, \citenamefont {Park}, \citenamefont {Lončar},\ and\
  \citenamefont {Lukin}}]{SR_SV_1_2016}%
  \BibitemOpen
  \bibfield  {author} {\bibinfo {author} {\bibfnamefont {A.}~\bibnamefont
  {Sipahigil}}, \bibinfo {author} {\bibfnamefont {R.~E.}\ \bibnamefont
  {Evans}}, \bibinfo {author} {\bibfnamefont {D.~D.}\ \bibnamefont {Sukachev}},
  \bibinfo {author} {\bibfnamefont {M.~J.}\ \bibnamefont {Burek}}, \bibinfo
  {author} {\bibfnamefont {J.}~\bibnamefont {Borregaard}}, \bibinfo {author}
  {\bibfnamefont {M.~K.}\ \bibnamefont {Bhaskar}}, \bibinfo {author}
  {\bibfnamefont {C.~T.}\ \bibnamefont {Nguyen}}, \bibinfo {author}
  {\bibfnamefont {J.~L.}\ \bibnamefont {Pacheco}}, \bibinfo {author}
  {\bibfnamefont {H.~A.}\ \bibnamefont {Atikian}}, \bibinfo {author}
  {\bibfnamefont {C.}~\bibnamefont {Meuwly}}, \bibinfo {author} {\bibfnamefont
  {R.~M.}\ \bibnamefont {Camacho}}, \bibinfo {author} {\bibfnamefont
  {F.}~\bibnamefont {Jelezko}}, \bibinfo {author} {\bibfnamefont
  {E.}~\bibnamefont {Bielejec}}, \bibinfo {author} {\bibfnamefont
  {H.}~\bibnamefont {Park}}, \bibinfo {author} {\bibfnamefont {M.}~\bibnamefont
  {Lončar}},\ and\ \bibinfo {author} {\bibfnamefont {M.~D.}\ \bibnamefont
  {Lukin}},\ }\bibfield  {title} {\bibinfo {title} {An integrated diamond
  nanophotonics platform for quantum-optical networks},\ }\href
  {https://doi.org/10.1126/science.aah6875} {\bibfield  {journal} {\bibinfo
  {journal} {Science}\ }\textbf {\bibinfo {volume} {354}},\ \bibinfo {pages}
  {847} (\bibinfo {year} {2016})}\BibitemShut {NoStop}%
\bibitem [{\citenamefont {Chu}\ \emph {et~al.}(2023)\citenamefont {Chu},
  \citenamefont {Papon}, \citenamefont {Bart}, \citenamefont {Wieck},
  \citenamefont {Ludwig}, \citenamefont {Midolo}, \citenamefont {Rotenberg},\
  and\ \citenamefont {Lodahl}}]{PhysRevLett.131.033606}%
  \BibitemOpen
  \bibfield  {author} {\bibinfo {author} {\bibfnamefont {X.-L.}\ \bibnamefont
  {Chu}}, \bibinfo {author} {\bibfnamefont {C.}~\bibnamefont {Papon}}, \bibinfo
  {author} {\bibfnamefont {N.}~\bibnamefont {Bart}}, \bibinfo {author}
  {\bibfnamefont {A.~D.}\ \bibnamefont {Wieck}}, \bibinfo {author}
  {\bibfnamefont {A.}~\bibnamefont {Ludwig}}, \bibinfo {author} {\bibfnamefont
  {L.}~\bibnamefont {Midolo}}, \bibinfo {author} {\bibfnamefont
  {N.}~\bibnamefont {Rotenberg}},\ and\ \bibinfo {author} {\bibfnamefont
  {P.}~\bibnamefont {Lodahl}},\ }\bibfield  {title} {\bibinfo {title}
  {Independent electrical control of two quantum dots coupled through a
  photonic-crystal waveguide},\ }\href
  {https://doi.org/10.1103/PhysRevLett.131.033606} {\bibfield  {journal}
  {\bibinfo  {journal} {Phys. Rev. Lett.}\ }\textbf {\bibinfo {volume} {131}},\
  \bibinfo {pages} {033606} (\bibinfo {year} {2023})}\BibitemShut {NoStop}%
\bibitem [{\citenamefont {Nguyen}\ \emph {et~al.}(2012)\citenamefont {Nguyen},
  \citenamefont {Sallen}, \citenamefont {Voisin}, \citenamefont {Roussignol},
  \citenamefont {Diederichs},\ and\ \citenamefont
  {Cassabois}}]{PhysRevLett.108.057401}%
  \BibitemOpen
  \bibfield  {author} {\bibinfo {author} {\bibfnamefont {H.~S.}\ \bibnamefont
  {Nguyen}}, \bibinfo {author} {\bibfnamefont {G.}~\bibnamefont {Sallen}},
  \bibinfo {author} {\bibfnamefont {C.}~\bibnamefont {Voisin}}, \bibinfo
  {author} {\bibfnamefont {P.}~\bibnamefont {Roussignol}}, \bibinfo {author}
  {\bibfnamefont {C.}~\bibnamefont {Diederichs}},\ and\ \bibinfo {author}
  {\bibfnamefont {G.}~\bibnamefont {Cassabois}},\ }\bibfield  {title} {\bibinfo
  {title} {Optically gated resonant emission of single quantum dots},\ }\href
  {https://doi.org/10.1103/PhysRevLett.108.057401} {\bibfield  {journal}
  {\bibinfo  {journal} {Phys. Rev. Lett.}\ }\textbf {\bibinfo {volume} {108}},\
  \bibinfo {pages} {057401} (\bibinfo {year} {2012})}\BibitemShut {NoStop}%
\bibitem [{\citenamefont {Nguyen}\ \emph {et~al.}(2013)\citenamefont {Nguyen},
  \citenamefont {Sallen}, \citenamefont {Abbarchi}, \citenamefont {Ferreira},
  \citenamefont {Voisin}, \citenamefont {Roussignol}, \citenamefont
  {Cassabois},\ and\ \citenamefont {Diederichs}}]{PhysRevB.87.115305}%
  \BibitemOpen
  \bibfield  {author} {\bibinfo {author} {\bibfnamefont {H.~S.}\ \bibnamefont
  {Nguyen}}, \bibinfo {author} {\bibfnamefont {G.}~\bibnamefont {Sallen}},
  \bibinfo {author} {\bibfnamefont {M.}~\bibnamefont {Abbarchi}}, \bibinfo
  {author} {\bibfnamefont {R.}~\bibnamefont {Ferreira}}, \bibinfo {author}
  {\bibfnamefont {C.}~\bibnamefont {Voisin}}, \bibinfo {author} {\bibfnamefont
  {P.}~\bibnamefont {Roussignol}}, \bibinfo {author} {\bibfnamefont
  {G.}~\bibnamefont {Cassabois}},\ and\ \bibinfo {author} {\bibfnamefont
  {C.}~\bibnamefont {Diederichs}},\ }\bibfield  {title} {\bibinfo {title}
  {Photoneutralization and slow capture of carriers in quantum dots probed by
  resonant excitation spectroscopy},\ }\href
  {https://doi.org/10.1103/PhysRevB.87.115305} {\bibfield  {journal} {\bibinfo
  {journal} {Phys. Rev. B}\ }\textbf {\bibinfo {volume} {87}},\ \bibinfo
  {pages} {115305} (\bibinfo {year} {2013})}\BibitemShut {NoStop}%
\bibitem [{SM(2023)}]{SM}%
  \BibitemOpen
  \href@noop {} {} (\bibinfo {year} {2023}),\ \bibinfo {note} {see
  Supplementary Material at \textit{inserturl} for a more detailed derivation
  of the analytical expressions as well as information about the numerically
  exact calculations.}\BibitemShut {Stop}%
\end{thebibliography}%
\end{document}